\documentclass[aps,twocolumn,nofootinbib,
preprintnumbers,superscriptaddress]{revtex4-1}

\usepackage{amsmath,amssymb,amsfonts,amsthm,graphicx, epsf,dsfont,dcolumn,yfonts}
\usepackage{latexsym}
\usepackage[colorlinks=true, pdfstartview=FitV, linkcolor=black, citecolor=black, urlcolor=black]{hyperref}

\usepackage{color}
\usepackage{slashed} 
\newcommand{\be}{\begin{equation}}
\newcommand{\ee}{\end{equation}}
\newcommand{\bea}{\begin{eqnarray}}
\newcommand{\eea}{\end{eqnarray}}
\def\ie{{\it i.e.~}}

\newcommand{\bwt}{\begin{widetext}}
\newcommand{\ewt}{\end{widetext}}

\def\G{\Gamma}

\def\l{\lambda}

\def\r{\rho}

\def\v{\varphi}

\begin{document}

\title{Pomeranchuk Instability in a non-Fermi Liquid from Holography
}
\author{Mohammad Edalati}
\affiliation{Weinberg Theory Group, University of Texas at Austin, Austin  TX 78712, USA}
\author{ Ka Wai Lo}
\author{Philip W. Phillips}
\affiliation{Department of Physics, University of Illinois at Urbana-Champaign, Urbana IL 61801, USA}
\vspace{15pt}

\preprint{UTTG-24-11}

\begin{abstract}

The Pomeranchuk instability, in which an isotropic Fermi surface distorts and becomes anisotropic due to strong interactions, is a possible mechanism for the growing number of experimental systems which display transport properties that differ along the $x$ and $y$ axes. We show here that the gauge-gravity duality can be used to describe such an instability in fermionic systems. Our holographic model consists of fermions in a background which describes the causal propagation of a massive neutral spin-two field in an asymptotically AdS spacetime. The Fermi surfaces in the boundary theory distort  spontaneously and become anisotropic once the neutral massive spin-two field develops a normalizable mode in the bulk. Analysis of the fermionic correlators reveals that the low-lying fermionic excitations are non-Fermi liquid-like both before and after the Fermi surface shape distortion.  Further, the spectral weight along the Fermi surface is angularly dependent and can be made to
vanish along certain directions. 

\end{abstract}

\maketitle

\section{Introduction}

\begin{figure}
\centering
\includegraphics[width=88mm]{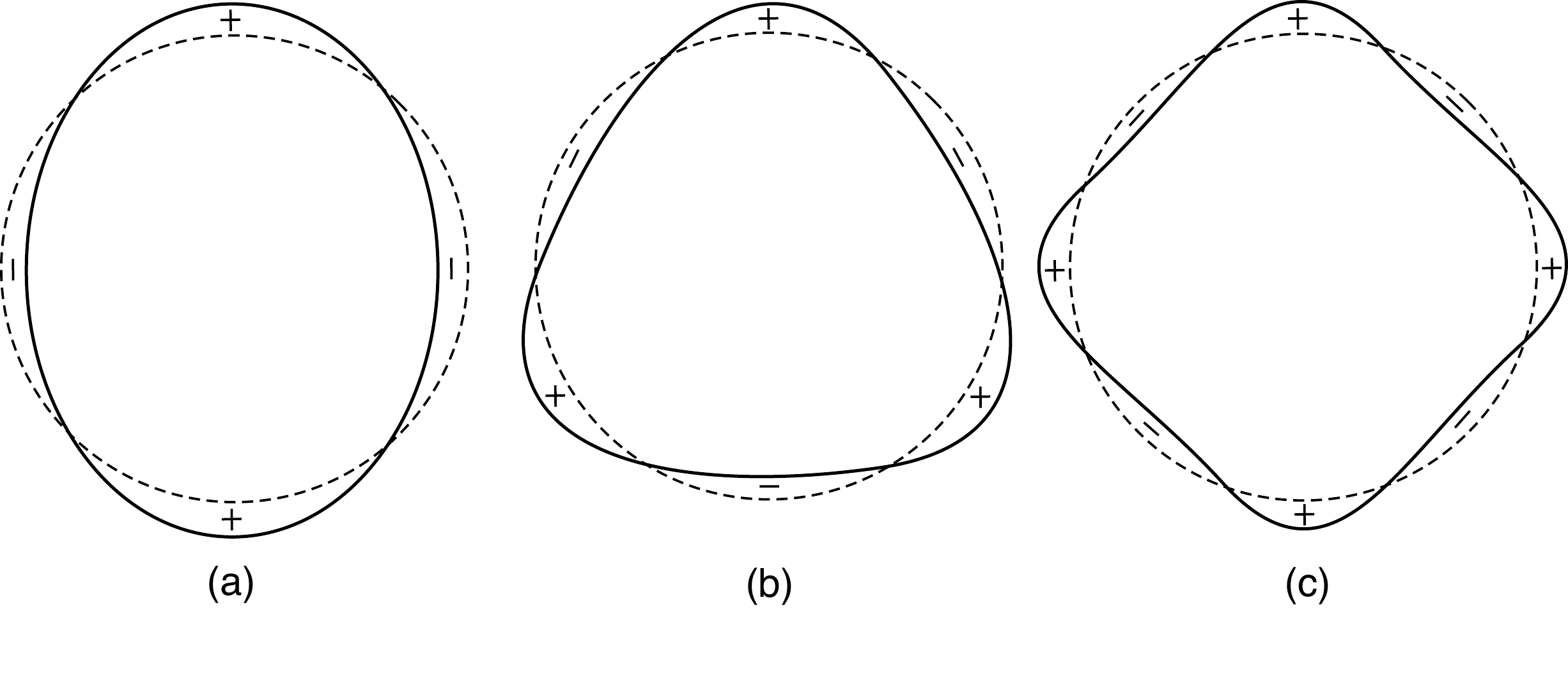}\vskip -0.1in
\caption{\label{Pomeranchuk Cartoon}\footnotesize{(a) Examples of the Pomeranchuk instability in (a) $l=2$, (b) $l=3$ and (c) $l=4$ channels. 
The $+$ and $-$ signs on the anisotropic Fermi surfaces show the momentum gain and loss compared to the isotropic Fermi surfaces (which are shown above by the dashed lines). } }
\end{figure}

To a surprising extent, the electronic properties of most metals are well described by Landau Fermi liquid theory. The key building blocks of this theory are quasi-particles and a Fermi surface, the surface in momentum space that demarcates filled from empty states. For a system of weakly interacting electrons, the shape of the Fermi surface is isotropic. On the other hand, if the interactions among the quasi-particles are strong enough, other possibilities might emerge.  Pomeranchuk \cite{Pomeranchuk1958} showed that  forward scattering among quasi-particles with non-trivial angular momentum and spin structure  can lead to an instability of the  ground state towards forming an anisotropic Fermi surface. For example, the simplest form of a two-dimensional nematic Fermi liquid can be thought of, in the continuum limit, as a realization of the Pomeranchuk instability in the angular momentum $l=2$ and spin $s=0$ particle-hole channel  whereby a circular Fermi surface spontaneously distorts and becomes elliptical with a symmetry of a quadrupole ($d$-wave). Such an instability, depicted in Figure \ref{Pomeranchuk Cartoon}(a),  is often referred to in the literature as the quadrupolar Pomeranchuk instability\footnote{The instabilities in other channels, including the antisymmetric spin channel, are also interesting. (Note that there is no Pomeranchuk instability in the $l=1$, $s=0$ channel.) In this paper, however, we only focus on the quadrupolar Pomeranchuk instability and the resulting nematic phase. }. In practice, however, the instability breaks the discrete point group symmetry of the underlying lattice. As such,  systems undergoing a Pomeranchuk instability exhibit manifestly distinct transport properties along the different crystal
axes.  Experimentally several instances of anisotropic transport in strongly correlated electronic systems have been reported \cite{Grigera:2004qe,Sun:2006,Ying:2011sp,Chu:2010dp} and occupy much of the current focus in correlated electron matter. Although it is unclear at present whether the Pomeranchuk mechanism is the only root cause\footnote{Nematic phases could also be obtained by the melting of stripe phases; see  \cite{FKELM2010,Fradkin2010} and references therein.}, 
it nevertheless offers a framework where such anisotropies can be characterized in detail.
So far, much of the work in this direction has focused on reaching the nematic phases of correlated electron matter, via the Pomeranchuk instability, in a weakly interacting Fermi fluid \cite{Halboth2000,Valenzuela2001}. A natural question to ask is whether a nematic phase could be approached, via a Pomeranchuk instability, from a non-Fermi liquid phase. To answer the question,  one has to address the origin of the Pomeranchuk instability beyond the Fermi liquid framework. Unlike Fermi liquids, there is no notion of stable quasi-particles in non-Fermi liquids. Nevertheless, non-Fermi liquids have sharp Fermi surfaces. Consequently,  a Pomeranchuk instability, viewed simply as a spontaneous shape distortion of the underlying Fermi surface, is still a theoretical possibility.

We explore here how such an instability can be studied using the
gauge-gravity duality, hereafter referred to as holography. The
duality, which maps certain $d$-dimensional strongly coupled quantum
field theories to ($d+1$)-dimensional semiclassical gravitational
theories in asymptotically AdS spacetimes, is a promising tool in
modeling some features of strongly correlated condensed matter
systems. For reviews on the applications of holography for condensed
matter physics, see \cite{Hartnoll:2009sz, Herzog:2009xv,
  McGreevy:2009xe, Horowitz:2010gk, Hartnoll:2011fn, Sachdev:2011wg,
  Iqbal:2011ae}. We take the system to be $d=2+1$ dimensional, as this
is the most interesting case experimentally
\cite{Grigera:2004qe,Sun:2006,Ying:2011sp,Chu:2010dp}. Also, we
consider the case where the broken phase is a nematic (non-)Fermi
liquid with the symmetry of a quadrupole. In other words, we only
explore the possibility of a quadrupolar Pomeranchuk instability. Such
an instability is characterized by an order parameter which is a
neutral symmetric traceless tensor ${\cal O}_{ij}$ where $i$ and $j$
denote spatial indices. Thus, in order to use holography to realize
the transition to such a phase in the boundary theory, one has to turn
on a neutral massive spin-two field $\varphi_{\mu\nu}$ in the bulk and
show that, under some circumstances, it develops a non-trivial
normalizable profile in the bulk radial direction. From the point of
view of the physics in the bulk, there is a serious hurdle that one
needs to overcome in order to successfully describe the propagation of
a massive spin-two field, which has to do with the existence of extra
unwanted degrees of freedom (ghosts) and superluminal
modes. Eliminating ghosts and superluminal modes within a theory that
also allows for the condensation of the operator dual to the massive
spin-two field is the first step. In addition, a coupling of the
correct type must be introduced in the bulk between the massive
spin-two field and the fermions  in order to bring about the shape
distortion of the Fermi surface in the boundary. We present here a
minimal model which can address each of these problem, and show that
the isotropic Fermi surface of the boundary non-Fermi liquid gets
spontaneously distorted and becomes elliptical. We also  find that the
broken (nematic) phase is a non-Fermi liquid.  We interpret this
transition as a holographic realization of the quadrupolar Pomeranchuk
instability in non-Fermi liquids.  Nonetheless, a disclaimer is
necessary here. The background we study cannot be extended
consistently to
$T=0$ on the count that so doing would require the back reaction of
the gauge field.  Such a back-reacted metric would be
non-Einstein and there is no causal Lagrangian for a spin-two
field in  a curved background that is not Einstein.  Within the
background we consider, the vaccuum
expectation value of the spin-two field
actually grows as the temperature is increased.  Consequently, there
is strictly no instability in the traditional sense in the context of
spontaneous symmetry breaking. Hence, our usage of the term
instability is not entirely warranted.  Nonetheless, the program we
outline here does provide a consistent holographic framework to address how to
engineer shape distortions of the underlying Fermi surface.  Hence, it
serves as a valid first step in realizing the physics of the
Pomeranchuk instability.

The outline of this paper is as follows.  Section II sets up the Lagrangian for the  neutral massive spin-two field.  Included here will be an explicit mechanism for obtaining a normalizable profile for the bulk spin-two field. Section III contains the fermonic degrees of freedom as well as the results illustrating that the low-lying excitations are non-Fermi liquid in nature as well as the shape distortion of the underlying Fermi surface.

\section{ Background}\label{section-setup}

Since we want to explore the possibility of a transition from a non-Fermi liquid to a nematic Fermi liquid  phase in the boundary, the first step is to construct a background which consistently incorporates a neutral massive spin-two field $\varphi_{\mu\nu}$. This field is dual to the operator whose vacuum expectation value (vev) is the order parameter for the nematic phase. 

\subsection{ BGP Lagrangian}

In flat  $(d+1)$-dimensional Minkowski spacetime, Fierz and Pauli suggested \cite{FP1939} a Lagrangian for a free neutral massive spin-two field which is quadratic in derivatives and correctly describes the propagating degrees of freedom for such a field.  Generalizing to curved spacetimes, Buchbinder, Gitman and Pershin \cite{Buchbinder:2000fy}, following \cite{Aragone:1971kh},  proposed  a simple Lagrangian (quadratic in the massive spin-two field, and also quadratic in derivatives) which describes the causal propagation of the correct number of degrees of freedom of a neutral massive spin-two field in a fixed Einstein background in arbitrary dimensions.  In four bulk dimensions, their Lagrangian for $\varphi_{\mu\nu}$, denoted by ${\cal L}_{\rm BGP}$, reads 
\begin{align}\label{BGPAction}
\frac{{\cal L}_{\rm BGP}}{\sqrt{-g}}=\frac{1}{4}&\Big\{-\nabla_\mu\varphi_{\nu\rho}\nabla^\mu\varphi^{\nu\rho}+\nabla_\mu\varphi\nabla^\mu\varphi\nonumber\\
&+2\nabla_\mu\varphi_{\nu\rho}\nabla^{\rho}\varphi^{\nu\mu}-2\nabla^\mu\varphi_{\mu\nu}\nabla^\nu\varphi\\
&-m^2\big(\varphi_{\mu\nu}\varphi^{\mu\nu}-\varphi^2\big)+\frac{R}{2}\, \varphi_{\mu\nu}\varphi^{\mu\nu}-\frac{R}{4}\, \varphi^2\Big\}.\nonumber
\end{align}
Here $R$ is the Ricci scalar. Also, we have defined $\varphi=\varphi^\mu_\mu$. Several comments are warranted here.  First, it is crucial that the fixed background satisfies the Einstein relation (which, in four bulk dimensions, reads) $R_{\mu\nu}= \Lambda g_{\mu\nu}$, with $\Lambda$  being the cosmological constant, otherwise some constraint equations become dynamical which would give rise to extra propagating degrees of freedom for  $\varphi_{\mu\nu}$. In our context this restriction simply implies that one should ignore the backreaction of $\varphi_{\mu\nu}$, as well as other bulk fields that will be introduced later, on the metric of the fixed Einstein background.  Second, in the flat spacetime limit, the above Lagrangian becomes the Fierz-Pauli Lagrangian \cite{FP1939}. Third, the massless limit of the BGP Lagrangian, \ie $m^2=0$, has an emergent gauge symmetry (diffeomorphism) which is unique to the graviton:  $\varphi_{\mu\nu}\rightarrow\varphi_{\mu\nu} +\nabla_{(\mu}\xi_{\nu)}$ with $\xi_\mu$ being an infinitesimal vector. Indeed, as pointed out in \cite{Buchbinder:2000fy}, this symmetry has been used to settle the ambiguity associated with the definition of the mass of $\varphi_{\mu\nu}$.  Fourth, for $m^2= R/6$ (here $d+1=4$) the above Lagrangian has another emergent gauge symmetry \cite{Deser:1983mm}:  $\varphi_{\mu\nu}\rightarrow\varphi_{\mu\nu}+(\nabla_\mu\nabla_\nu+g_{\mu\nu}R/12)\epsilon$ with $\epsilon$ being an infinitesimal scalar gauge parameter. In the following, we always assume that $m^2\neq 0$ and $m^2\neq R/6$. Excluding these two special values for the mass, the equations of motion and constraints obtained from \eqref{BGPAction} are given by \cite{Buchbinder:2000fy}
\begin{align}\label{DynamicEOM}
0=&\left(\nabla^2-m^2\right)\varphi_{\mu\nu}+2R^{\rho\,\,\,\sigma}_{\,\,\,\mu\,\,\,\nu}\varphi_{\rho\sigma},\\
0=&\nabla^\mu\varphi_{\mu\nu},\\
0=&\varphi,\\
0=&{\dot\varphi},\\
0=&g^{00}\nabla_0\nabla_i\varphi^i_{\,\nu}-g^{0i}\nabla_0\nabla_i\varphi^0_{\,\nu}-g^{0i}\nabla_i\nabla_0\varphi^0_{\,\nu}\nonumber\\
&-g^{ij}\nabla_i\nabla_j\varphi^0_{\,\nu}-2R^{\rho\,0 \,\sigma}_{\,\,\,\,\,\,\,\,\,\,\,\nu}\,\varphi_{\rho\sigma}+m^2\varphi^{0}_{\,\nu}.\label{Const4Eq}
\end{align}
The above expressions give  the correct number of propagating degrees of freedom for a massive spin-two field in (3+1)-dimensions. Such a field transforms in the five-dimensional irreducible representation of SO(3), with SO(3) being the little group of the Lorentz group SO(3,1).

As we alluded to earlier, in order to satisfy the constraint  $R_{\mu\nu}= \Lambda g_{\mu\nu}$, we will work in a regime where matter fields do not backreact on the bulk geometry. So, we take the Schwarzschild AdS$_4$ black hole as our background geometry ($\Lambda =-3$)

\begin{align}\label{SchMetric}
ds^2&=g_{\mu\nu}dx^\mu dx^\nu\nonumber\\
&=r^2\left(-f(r)\,dt^2+dx^2+dy^2\right)+\frac{dr^2}{r^2f(r)},
\end{align}
where $f(r)=1-(r_0^3/r^3)$ with $r_0$ being the horizon radius, given by the largest real root of $f(r_0)=0$.
Note that we are working in units where we set the curvature radius $L$ of AdS$_4$ equal to unity.    Also, note that in the coordinates we have chosen above, the asymptotic boundary of the spacetime is at $r\to\infty$. The temperature of the black hole is given by $T=3r_0/(4\pi)$.

To solve for $\varphi_{\mu\nu}$, we only consider the configuration where $\varphi_{\mu\nu}=\varphi_{\mu\nu}(r)$ for all $\mu,\nu\in \{t,r,x,y\}$. Analyzing the equations \eqref{DynamicEOM}--\eqref{Const4Eq},  one can show \cite{Benini:2010qc, Benini:2010pr} that it is consistent to put $\varphi_{\mu\nu}(r)$ in the form 
\begin{align}\label{SpinTwoAnsatz}
\hskip-0.05in\varphi_{\mu\nu}(r)=\left(
\begin{array}{cc}
\mathbf{0}  & \mathbf{0}     \\
\mathbf{0}  &  \mathbf{\varphi}_{ij}(r)   
\end{array}
\right), 
\end{align}
where $\varphi_{ij}(r)$, with $i,j=x,y$, is a symmetric traceless two-by-two matrix.
Defining the ``director" $\vec\varphi(r)= \varphi_{xx}(r)+i\varphi_{xy}(r)$,  under a rotation $\theta$ in the ($x$,$y$)-plane, $\vec\varphi(r)$ transforms as
\begin{align}\label{GaugeChoice}
\vec\varphi(r) \to e^{2i\theta} \vec\varphi(r).
\end{align}
One can then choose a particular value of $\theta$ to set either $\varphi_{xx}(r)$ or  $\varphi_{xy}(r)$ equal to zero, or make them equal. 

Suppose now there exists a (2+1)-dimensional field theory dual to the gravitational setup under consideration here. Focusing for the moment on the bulk neutral massive spin-two field $\varphi_{\mu\nu}$, holography tells us that it should be dual to some operator in the boundary theory. Since $\varphi_{\mu\nu}$ is a bulk field with spin greater than zero, its number of components does not naively match the number of components of the dual operator. But, as explained in \cite{Benini:2010pr}, since $\varphi_{\mu\nu}$ can be put in the form \eqref{SpinTwoAnsatz}, it is $\varphi_{ij}$ which sources a neutral symmetric traceless operator ${\cal O}_{ij}$ in the boundary theory. 
(The vev of the operator ${\cal O}_{ij}$ determines the nematicity of the (2+1)-dimensional boundary theory.) More concretely, equation \eqref{DynamicEOM}  implies that $\varphi_{ij}(r)$ takes the following form near the boundary as $r\to\infty$
\begin{align}\label{AsymSpinTwo}
\varphi_{ij}(r)=A_{ij}\,r^{\Delta-1}\big(1+\cdots\big)+B_{ij}\,r^{2-\Delta}\big(1+\cdots\big),
\end{align}
with $A_{ij}$ and $B_{ij}$ being symmetric traceless tensors. $A_{ij}$ is the source for the boundary theory operator  ${\cal O}_{ij}$ while $B_{ij}$ is proportional to its vev, $\langle{\cal O}_{ij}\rangle$.  In \eqref{AsymSpinTwo}, $\Delta$ denotes  the UV scaling dimension of  ${\cal O}_{ij}$, which is related to the mass of the bulk field $\varphi_{\mu\nu}$ via 
\begin{align}\label{Deltavsmass}
\Delta=\frac{3}{2}+\sqrt{\frac{9}{4}+m^2}.
\end{align}
Note that for $m=0$, \ie when $\varphi_{\mu\nu}$ is the graviton, equation \eqref{Deltavsmass} yields $\Delta=3$ which is the dimension of $T_{\mu\nu}$, the energy-momentum tensor operator,  of the (2+1)-dimensional boundary theory. 
There is a  Breitenlohner-Freedman bound \cite{Breitenlohner:1982bm} for the propagation of $\varphi_{\mu\nu}$ in an asymptotically AdS$_4$ geometry which reads $m^2\geq 0$ \cite{Deser:2001us, Deser:2001wx, Benini:2010pr}. Equation \eqref{Deltavsmass} then implies that  $\Delta\geq 3$.
In our discussions in this paper, we will always take the mass squared of $\varphi_{\mu\nu}$ in the asymptotic AdS$_4$ region of the geometry to satisfy the condition $m^2>0$. Note that the condition $m^2\neq R/6=-2$ is then trivially satisfied. 

Given the ansatz \eqref{SpinTwoAnsatz}, we are interested in solutions in which $\varphi_{\mu\nu}(r)$, or rather $\varphi_{ij}(r)$, is regular near the horizon and normalizable as $r\to\infty$. More concretely, near the horizon, we look for a solution of the equation \eqref{DynamicEOM}  which is regular as  $r\to r_0$, namely 
\begin{align}\label{HorizonBehavior}
\varphi_{ij}(r)=a_{ij} +b_{ij}(r-r_0)+c_{ij} (r-r_0)^2+\cdots, 
\end{align}
where the coefficient tensors $b_{ij}$, $c_{ij}$, $\cdots$ are all determined in terms of $a_{ij}$. In the boundary as $r\to \infty$, we demand $\varphi_{ij}(r)$ to be normalizable
\begin{align}\label{BoundaryBehavior}
\varphi_{ij}(r)=B_{ij}\,r^{2-\Delta}\big(1+\cdots\big). 
\end{align}
If a solution with the above two boundary conditions exists, then the boundary theory would be in a phase where the neutral symmetric traceless operator ${\cal O}_{ij}$ (which is the operator dual to $\varphi_{ij}$) spontaneously condenses.  It is easy to show that given the set of equations \eqref{DynamicEOM}--\eqref{Const4Eq}, such a solution does not exist. As we will explain in the next section, one way forward is to minimally modify the BGP Lagrangian \eqref{BGPAction} such that the new action yields the correct number of causal propagating degrees of freedom for $\varphi_{\mu\nu}$, and also permits the aforementioned spontaneous condensation to occur. 

\subsection{Modifying the BGP Lagrangian}

Undoubtedly, there are many ways to modify the BGP Lagrangian to
facilitate  spontaneous condensation of the operator dual to
$\varphi_{\mu\nu}$.   Instead of categorizing all such modifications,
we take perhaps the simplest possibility.  Gubser \cite{Gubser:2005ih} has shown that
coupling a neutral scalar field to the square of the Weyl tensor leads
to a scalar hair on an asymptotically flat Schwarzschild black
hole. This coupling is of interest here because the square of the Weyl
tensor vanishes at the boundary ($r\rightarrow\infty$) and hence does
not affect the scaling dimension of the dual boundary operator.   To this end, we 
modify the BGP Lagrangian \eqref{BGPAction} as follows
\begin{align}\label{CondensateLagrangian}
\hskip-0.05in{\cal L}_{\rm \varphi}={\cal L}_{\rm BGP}+\frac{\ell^2}{4}\sqrt{-g}\,C_{\mu\nu\rho\sigma} C^{\mu\nu\rho\sigma}\left(\varphi_{\gamma\delta}\varphi^{\gamma\delta}-\varphi^2\right),
\end{align}
with $C_{\mu\nu\rho\sigma}$ the Weyl tensor.
 In the Appendix, we show that  the above modification  to the BGP Lagrangian, though not unique, offers a valid description of the neutral massive spin-two field $\varphi_{\mu\nu}$, meaning that $\varphi_{\mu\nu}$ still propagates causally in a fixed Einstein spacetime with the correct number of degrees of freedom. 
In what follows, we show that the equations of motion for $\varphi_{\mu\nu}$ obtained from the new Lagrangian now admits non-trivial normalizable solutions (whose near horizon and asymptotic boundary behaviors are given by \eqref{HorizonBehavior} and \eqref{BoundaryBehavior}, respectively). 

\subsection{Normalizable Solution}

From the Appendix,  the relevant equation for determining whether there exists a normalizable solution for $\varphi_{\mu\nu}$ is
\begin{align}\label{ModifiedDynamicEOM}
\left(\nabla^2-m^2+\ell^2 C_{\gamma\delta\rho\sigma} C^{\gamma\delta\rho\sigma}\right)\varphi_{\mu\nu}+2R^{\rho\,\,\,\sigma}_{\,\,\,\mu\,\,\,\nu}\varphi_{\rho\sigma}=0.
\end{align}
Note that the background metric must again satisfy the Einstein relation. Since we have taken the spacetime to be the Schwarzschild AdS$_4$ black hole,  the square of the Weyl tensor is $C_{\mu\nu\rho\sigma} C^{\mu\nu\rho\sigma}= 12(r_0/r)^6$. 
One can easily show that, given the constraints presented in the Appendix and the equation \eqref{ModifiedDynamicEOM}, it is again consistent to put $\varphi_{\mu\nu}$ in the form given in \eqref{SpinTwoAnsatz}. Notice that as shown above, the square of the Weyl tensor evaluated on the background vanishes as $r\to\infty$. This then implies that the relationship between the asymptotic mass $m$ of  $\varphi_{ij}$ and the UV dimension $\Delta$ of the dual operator ${\cal O}_{ij}$ is still given by \eqref{Deltavsmass}. Also,  the mass of $\varphi_{ij}$ in the asymptotic AdS$_4$ region satisfies the same Breitenlohner-Freedman bound as before. Thus, regardless of the value of the coupling $\ell^2$, the asymptotic AdS$_4$ region of the background remains stable. 

\begin{figure}
\centering
\hskip -0.09in\includegraphics[width=8.0cm]{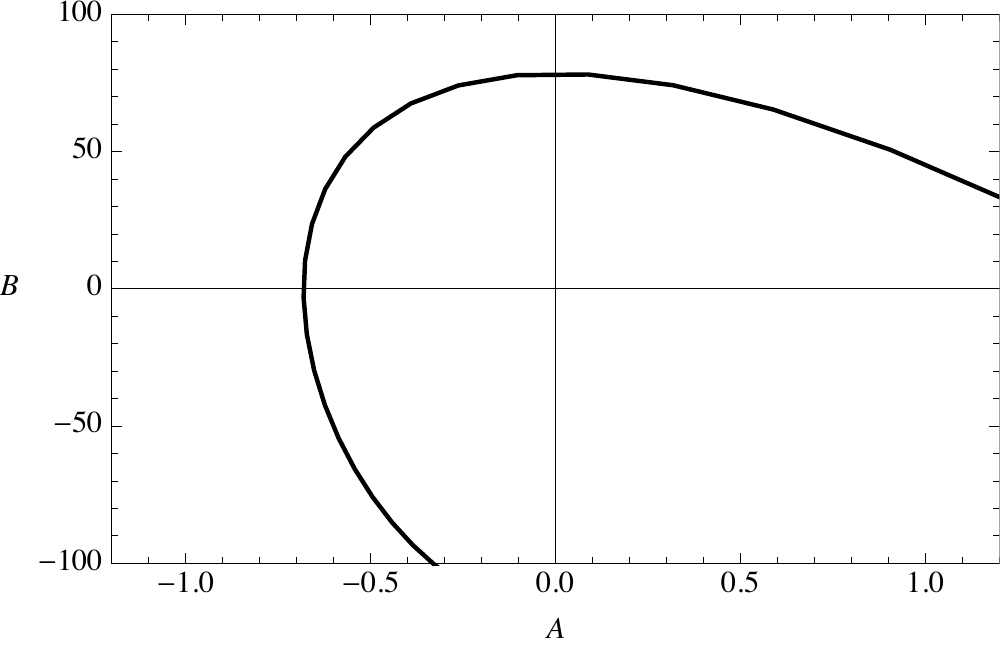}
\caption{\label{conden}\footnotesize{A plot of $B_{ij}$ verse $A_{ij}$ for $\ell^2=1.0866$ and $m^2=1/100$.   The values of  $B_{ij}$ and $A_{ij}$ have been rescaled so that they are dimensionless. The gauge in \eqref {GaugeChoice} has been used to set $\varphi_{xx}=\varphi_{xy}$. } }
\end{figure}

As Figure \ref{conden} illustrates, there exists a normalizable solution with 
$A_{ij}=0$ and $B_{ij}\ne0$. (We have used the gauge in \eqref {GaugeChoice} to set $\varphi_{xx}=\varphi_{xy}$.)  The holographic interpretation of this normalizable solution is that  ${\cal O}_{ij}$ has spontaneously condensed in the boundary theory.  
While it may be
possible to obtain a normalizable solution for $\varphi_{\mu\nu}$ using  an alternative mechanism,  we believe that the details of how the dual operator condenses in the boundary theory should be irrelevant to the Pomeranchuk instability as we discuss in the next section.

\subsection{Maxwell's Lagrangian}

Since we will be considering the boundary theory  at a finite chemical potential $\mu$ for a U(1) charge, we need to introduce in the bulk an Abelian gauge field $A_{\mu}$ with the Lagrangian
\begin{align}\label{MaxwellAction}
{\cal L}_{\rm M} = -\frac{1}{4}\sqrt{-g}\,F_{\mu\nu}F^{\mu\nu}. 
\end{align}
The equations of motion for $A_\mu$ is then
\begin{align}\label{MaxwellEOM}
\nabla^\mu F_{\mu\nu}=0.
\end{align}
We will only be interested in the case where $A_t$ is non-zero and, moreover, depends only on the radial coordinate $r$. So, we take the ansatz $A_{\mu}=A_t(r)\delta_{\mu0}$. Given our assumption that $A_{\mu}$ does not back react on the metric \eqref{SchMetric}, the solution to \eqref{MaxwellEOM} is easily obtained to be
\begin{align}\label{GaugeField}
A=\mu\Big(1-\frac{r_0}{r}\Big)dt,
\end{align}
where the boundary conditions at the horizon and the asymptotic boundary have been fixed by demanding $A_t(r_0)=0$ and $A_t(r\to\infty)=\mu$.

\section{Fermions}

In this section, we study the shape distortion of the Fermi surface as a result of the condensation of the boundary theory operator ${\cal O}_{ij}$.  We start by introducing a bulk spinor field $\psi$ with mass $m_\psi$ and charge $q$ which is dual to a boundary theory fermionic operator ${\cal O}_{\psi}$ with a UV scaling dimension $\Delta_\psi$ and charge $q$.  Note that the bulk spinor $\psi$ is a four-component Dirac spinor while the dual operator ${\cal O}_\psi$ is a two-component Dirac spinor. In this paper, we take $m_\psi\in[0, \frac{1}{2})$ and choose  the so-called ``conventional quantization" where the scaling dimension of the fermionic operator is related to the mass of the bulk spinor field through $\Delta_\psi=\frac{3}{2}+m_\psi$. 

\subsection{Fermion Lagrangian and Equation of Motion}

In our discussions below, the bulk spinor field $\psi$ is treated as a probe where its backreaction on the background metric, gauge field and neutral massive spin-two field is ignored. The Dirac Lagrangian,
\begin{align}\label{DiracLagrangian}
{\cal L}_{\rm Dirac}=-\sqrt{-g}\,i\bar\psi (\slashed{D}-m)\psi,
\end{align}
describes the bulk spinor field, where $\bar\psi=\psi^\dagger\,\Gamma^{\underline t}$ and $\slashed{D}=e_c^\mu\G^c\left(\partial_\mu+\frac{1}{4}\omega_\mu^{~ab}\G_{ab}-iqA_\mu\right)$
and $e^\mu_a$ and $\omega_\mu^{~ab}$ being respectively the vielbein and the spin connection. Also,  $a,b,\cdots=\{\underline{t},  {\underline x}, {\underline y}, \underline r\}$ denote the tangent space indices, $\Gamma^{\underline t},\Gamma^{\underline x}, \Gamma^{\underline y},\Gamma^{\underline r}$ are the Dirac matrices satisfying the Clifford algebra $\{\Gamma^a,\Gamma^b\}=2\eta^{ab}$ and  $\G_{ab}=\frac{1}{2}[\Gamma_a, \Gamma_b]$.  The overall negative sign in \eqref{DiracLagrangian}  ensures that the bulk action, once holographically renormalized, gives rise to a positive spectral density for the fermionic operator ${\cal O}_{\psi}$. In \eqref{DiracLagrangian},  $g_{\mu\nu}$ and $A_\mu$ are the background metric and the background gauge field whose expressions are given in \eqref{SchMetric} and \eqref{GaugeField}, respectively. 

In the probe limit that we are considering here the Dirac Lagrangian \eqref{DiracLagrangian} is not capable of holographically realizing how the spectral function of the boundary theory fermionic operator ${\cal O}_{\psi}$ is affected in the presence the order parameter $\langle {\cal O}_{ij}\rangle$.
To do so, one has to explicitly introduce bulk couplings between the
spinor field $\psi$ and  the neutral massive spin-two field
$\varphi_{\mu\nu}$. A similar problem arises in the context of
holographic $d$-wave superconductivity \cite{Benini:2010qc,
  Benini:2010pr} although in that case one has to deal with coupling
the bulk spinor to a charged massive spin-two field.  Since our goal
here is to capture the leading order effects on the retarded two-point
function of the fermionic operator due to the presence of the order
parameter, we consider those couplings in the bulk which are quadratic
in the spinor field and have relatively low mass
dimension. Furthermore, among the aforementioned  couplings, we will
only be interested in those which will potentially give rise to
anisotropic features in the fermion spectral function. In other words,
we ignore the terms which would just modify the mass of the spinor
field. Leaving aside those terms which would vanish once evaluated on
the background, we focus on the leading-order\footnote{There is also a
  coupling of the form $\lambda_5\sqrt{-g}\,\varphi_{\mu\nu}
  \bar\psi\, \Gamma^5\Gamma^\mu D^\nu\psi$ that breaks parity in the
  boundary theory when the operator dual to $\varphi_{\mu\nu}$
  condenses. We will not consider such a coupling in this paper. 
}  non-trivial coupling (which can potentially contribute asymmetric features to the fermionic spectral function) 
\begin{align}
{\cal L}_{\rm int}=-i\lambda\sqrt{-g} \,\varphi_{\mu\nu} \bar\psi\,\Gamma^\mu D^\nu \psi, 
\end{align}
with $\lambda$ real. Thus, the bulk fermion Lagrangian that we study is ${\cal L}_{\psi}={\cal L}_{\rm Dirac}+{\cal L}_{\rm int}$.  

The Dirac equation following from ${\cal L}_{\psi}$ takes the form
\begin{align}\label{DiracEquation}
(\slashed{D}-m+\lambda \varphi_{\mu\nu}\Gamma^\mu D^\nu)\psi =0.
\end{align}
To solve the above equation, we go to momentum space by Fourier transforming $\psi(r,x^\mu) \sim e^{ik.x}\psi(r,k^\mu)$, where $k^\mu = (\omega,\vec k)$. It is expedient to choose a basis for the Dirac matrices as follows
\begin{align}\label{GammaBasis}
 \Gamma^{\underline r}-- &= \begin{pmatrix}
   -\sigma_3  & 0 \\
    0&   -\sigma_3 
   \end{pmatrix}, \hskip 0.2in\Gamma^{\underline t} = \begin{pmatrix}
   i\sigma_1 & \,\,\,\,0 \\
    0&   \,\,\,\,i\sigma_1 
   \end{pmatrix},\nonumber\\ \Gamma^{\underline x} &= \begin{pmatrix}
   -\sigma_2  & \,\,\,\,0 \\
    0&   \,\,\,\,\sigma_2 
   \end{pmatrix}, \hskip 0.2in \Gamma^{\underline y} = \begin{pmatrix}
  \, 0  & \,\,\,\,\,\,\sigma_2 \\
    \,\,\sigma_2&  \,\,\,\,\,0 
   \end{pmatrix}.
   \end{align}
It is also convenient to remove the spin connection from the Dirac operator $\slashed{D}$ by rescaling $\psi$ according to $\psi\to(-g g^{rr})^{-1/4}\psi$. Splitting the new (rescaled) spinor $\psi$ according to $\psi^T=(\psi_1,\psi_2)$, where $\psi_1(r; \omega, \vec k)$ and $\psi_2(r; \omega, \vec k)$ are two-component spinors, and given the ansatz for the background form of $\varphi_{\mu\nu}$ where the only non-vanishing components are $\varphi_{xx}=-\varphi_{yy}$ and $\varphi_{xy}=\varphi_{yx}$, the equation \eqref{DiracEquation} results in the following two coupled differential equations for $\psi_1$ and $\psi_2$
\begin{align}\label{CoupledDiracEquation1}
\left({\tilde D}_r-i\sigma_2\sqrt{g^{xx}}\,{\tilde k}_x\right)\psi_1+i\sigma_2\sqrt{g^{xx}}\,{\tilde k}_y \psi_2&=0,\\
\left({\tilde D}_r+i\sigma_2\sqrt{g^{xx}}\,{\tilde k}_x\right)\psi_2+i\sigma_2\sqrt{g^{xx}}\,{\tilde k}_y \psi_1&=0,\label{CoupledDiracEquation2}
\end{align}
where we have defined 
\begin{align}\label{tildeD}
{\tilde D}_r&=-\sigma_3\sqrt{g^{rr}}\,\partial_r+\sigma_1\sqrt{-g^{tt}}\left(\omega+qA_2\right)-m_\psi,\\
{\tilde k}_x&=k_x\left(1+\lambda g^{xx}\varphi_{xx}\right)+\lambda g^{xx}\varphi_{xy} k_y,\label{tildekx}\\
{\tilde k}_y&=k_y\left(1-\lambda g^{xx}\varphi_{xx}\right)+\lambda g^{xx}\varphi_{xy} k_x.\label{tildeky}
\end{align} 
For $\lambda=0$, or when $\varphi_{xx}=\varphi_{xy} = 0$, the Dirac equation \eqref{DiracEquation} has been studied extensively in the literature (although mainly on the Reissner-Nordstr\"om AdS$_{4}$  black hole background), following the work of \cite{Lee:2008xf, Liu:2009dm, Cubrovic:2009ye, Faulkner:2009wj}. These studies show that in these cases there are symmetrical Fermi surfaces in the boundary theory whose underlying excitations can either be Fermi or non-Fermi liquid-like. On the other hand, when $\varphi_{xx}$ develops a normalizable mode in the bulk, $\tilde k_x$ and $\tilde k_y$ will differ thereby giving rise to asymmetrical features in the fermionic correlators of the boundary theory.  This can be shown explicitly.

To compute the retarded correlator of the boundary theory operator ${\cal O}_\psi$, one needs to solve the equations \eqref{CoupledDiracEquation1} and \eqref{CoupledDiracEquation2} with in-falling boundary condition at the horizon \cite{Son:2002sd} and read off the source and the expectation value of ${\cal O}_\psi$ from the asymptotic expansion of $\psi_\alpha$ ($\alpha =1,2$) following the prescription of \cite{Iqbal:2009fd}. Indeed, choosing in-falling boundary conditions for $\psi_\alpha$ near the horizon,  the leading-order asymptotic ($r\to\infty$) behavior of $\psi_\alpha$  takes the form $\psi_\alpha(r\to\infty)=(B_\alpha\, r^{-m_\psi}, A_\alpha r^{m_\psi})^T$, where the two-component spinor $A=(A_1, A_2)^T$ sources the boundary theory fermionic operator ${\cal O}_{\psi}$, while $B=(B_1, B_2)^T$ gives the vev of the operator. The spinor $B$ is related to the spinor $A$ through $B={\cal S}A$ from which the retarded Green function of the operator ${\cal O}_{\psi}$ take the form \cite{Iqbal:2009fd}  
\begin{align}\label{GreenFunction}
G_R(\omega,\vec k)=-i {\cal S}\gamma^t.
\end{align}
Note that in our basis of gamma matrices  $\gamma^t=i\sigma_1$. Up to a numerical constant, the spectral function of  the operator ${\cal O}_{\psi}$ is given by  $\rho(\omega,\vec k)={\rm Tr\,Im} G_R(\omega,\vec k)$.

\subsection{Fermi Surfaces and Low-Energy Excitations}

\begin{figure}
\centering
\hskip -0.225in\includegraphics[width=92mm]{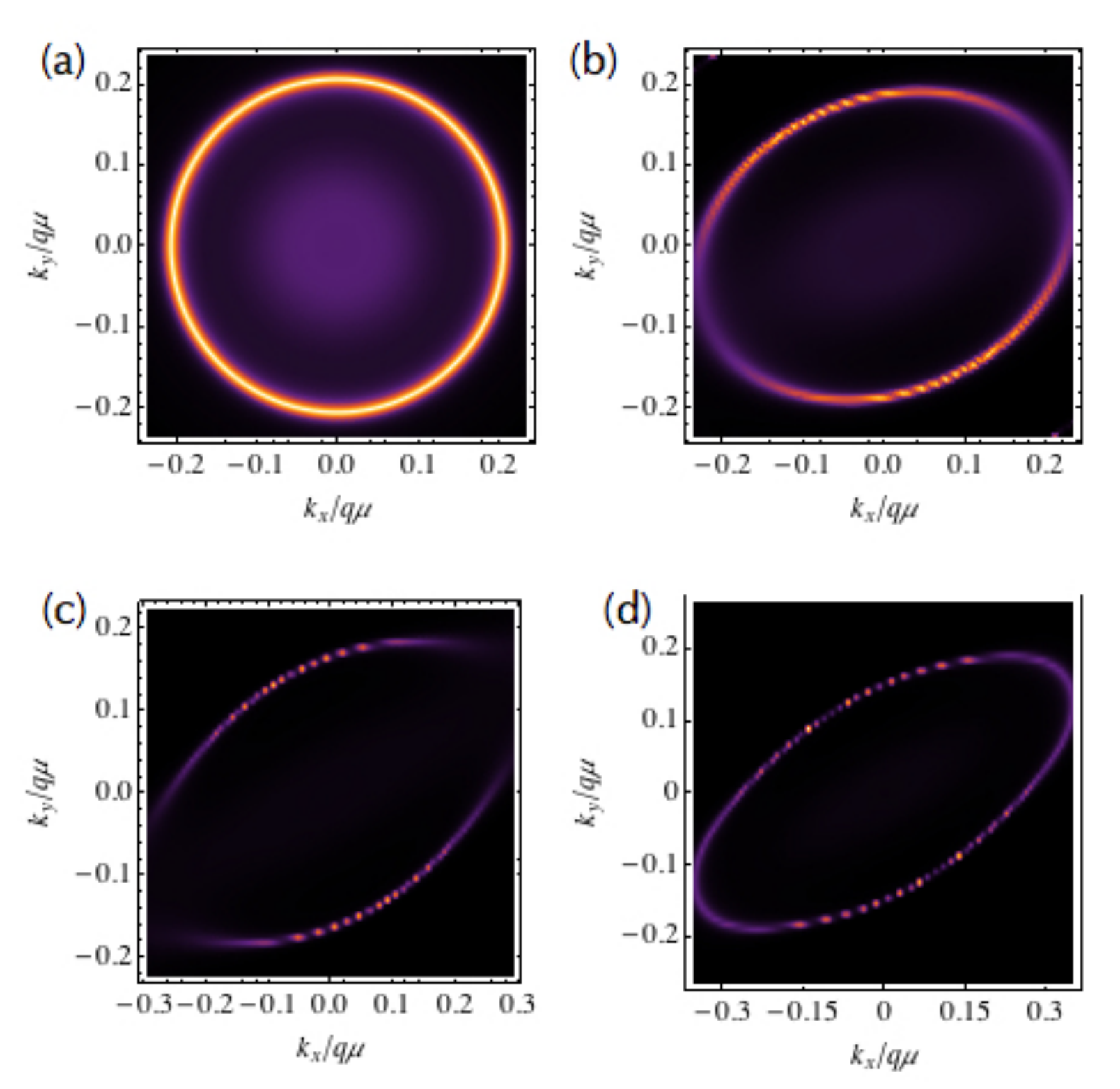}
\caption{\label{po} \footnotesize{Density plots of the spectral function of
the fermionic operator ${\cal O}_{\psi}$ in the boundary field theory. a) Spectral density when the boundary field theory is in the unbroken phase b) Spectral density in the broken phase showing an elliptic Fermi surface where we set $\l=-0.4$,  c) spectral density in the broken phase where $\l=-0.95$ showing a suppression of the spectral weight and d) spectral density, also in the broken phase, when $\l=-1.4$. } }
\end{figure}

To obtain the retarded Green function \eqref{GreenFunction} of the fermionic operator ${\cal O}_\psi$, one should numerically solve the equations \eqref{CoupledDiracEquation1} and \eqref{CoupledDiracEquation2}, or the flow equations obtained from them, with in-falling boundary condition at the horizon. We do this computation at finite temperature mainly because our background, in which there is a normalizable solution for the neutral massive spin-two field, is not,  strictly speaking, valid at zero temperature. The Femi momentum $\vec k_{\rm F}$ appears as a pole in ${\rm Re}\,G_R(\omega,\vec k)$, although at finite temperature such poles, instead of being the delta function, become broadened.  

Carrying out the numerical computation, we find multiple Fermi surfaces for the parameters chosen. To demonstrate proof of concept, it is sufficient to focus on the first Fermi surface, where $|\vec k_{\rm F}|$ (measured compared to the effective chemical potential $q\mu$) has the smallest value. In generating the plots shown in this section, we first used the gauge \eqref{GaugeChoice} to set $\varphi_{xx}=\varphi_{xy}$ and then numerically solved the Dirac equations for a very small but non-zero frequency such that the delta functions at the Fermi surface are broadened. 

\begin{figure}
\centering
\hskip -0.2in\includegraphics[width=85mm]{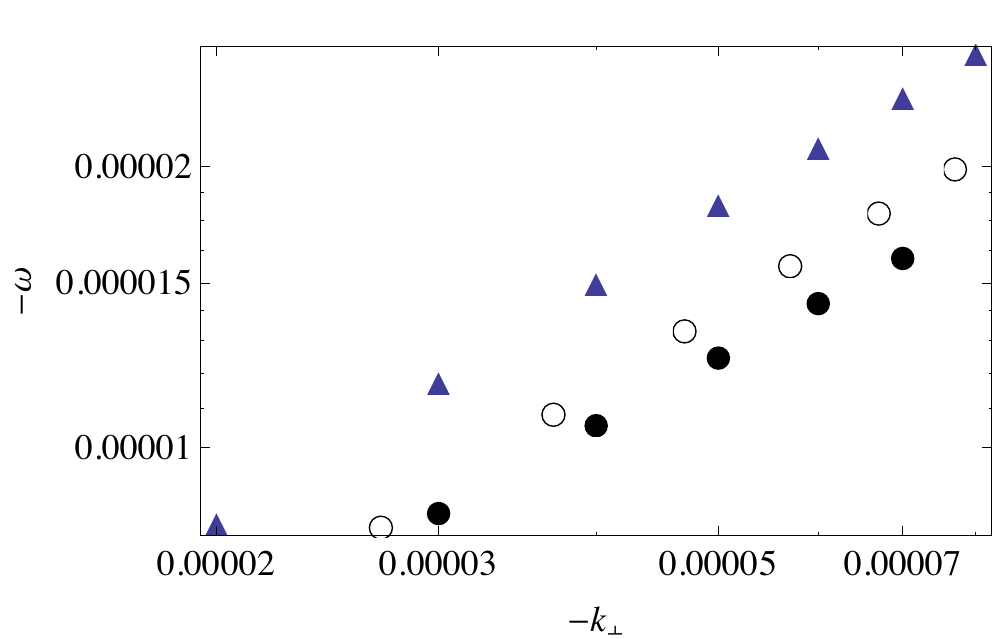}
\caption{\label{kperp} \footnotesize{The dispersion relation, $\omega(k)\propto k_\perp^z$, of quasi-particles around $k_{\rm F}$ plotted on a log-log scale.  The three cases correspond to 
 a) open circles for the unbroken phase of the boundary theory with a circular Fermi surface where we find  $z=0.64$, b) solid circles for the dispersion relation (in the broken phase) along
 $x$-axis where we find $z=0.74$ with $\lambda=-0.4$ and c) solid triangles for the dispersion relation along
 $y$-axis (in the broken phase) where $z=0.84$ with $\lambda =-0.4$.} }
\end{figure}

First, consider the case where the background value of $\varphi_{\mu\nu}(r)$ identically vanishes in the bulk. In other words the boundary theory is in the symmetry-unbroken phase $\langle {\cal O}_{ij}\rangle=0$. Not surprisingly, the Fermi surface obtained in this phase is isotropic, as is evident from Figure \ref{po}(a). To investigate the nature of the excitations, we computed the quasiparticle dispersion relation by focusing on the peak in the spectral function. In a Fermi liquid, $\omega(k)= k_\perp^z$ with $k_\perp= k-k_{\rm F}$ and $z=1$.   Figure \ref{kperp} (see open circles) reveals that in the unbroken phase of the boundary theory where the Fermi surface is circular, the dynamical exponent $z=0.64$ which indicates that the low-lying excitations form a non-Fermi liquid.  Importantly then, any subsequent breaking of rotational symmetry of the Fermi surface will be from a non-Fermi liquid state, one of the key hurdles in describing the strong correlations in the experimental systems.

Focusing now on the interesting phase where the neutral symmetric traceless operator ${\cal O}_{ij}$ has spontaneously condensed, we find three significant features in the spectral function of the fermionic operator ${\cal O}_{\psi}$.   First, the rotational symmetry of the Fermi surface is broken. As we alluded to earlier, such behavior is expected given the form of ${\tilde k}_x$
and ${\tilde k}_y$ at each $r$-slicing in the bulk.  Second,  the
spectral weight is angularly dependent.  For intermediate values of
the coupling $\lambda$, a gap-like feature opens at the two end points of the
major axis. This highly non-trivial behavior is not a generic feature
of traditional treatments of the Pomeranchuk instability
\cite{Fradkin2010} but can be understood
within a WKB approximation to the spectral function as outlined in the
next subsection. The diminished spectral density reappears once $\lambda$ exceeds a critical
value as shown in Figure \ref{po}(d).   
Finally, the excitations near the Fermi surface remain non-Fermi liquid-like.  The solid circles and triangles in Figure \ref{kperp} show that the dispersions along $k_x$ and $k_y$
deviate from linearity with exponents of $z=0.74$ and $z=0.84$,
respectively.  These values are, of course, parameter dependent.

\subsection{WKB Analysis of Spectral Function}

\begin{figure}
\hskip -0.09in\includegraphics[width=8.0cm]{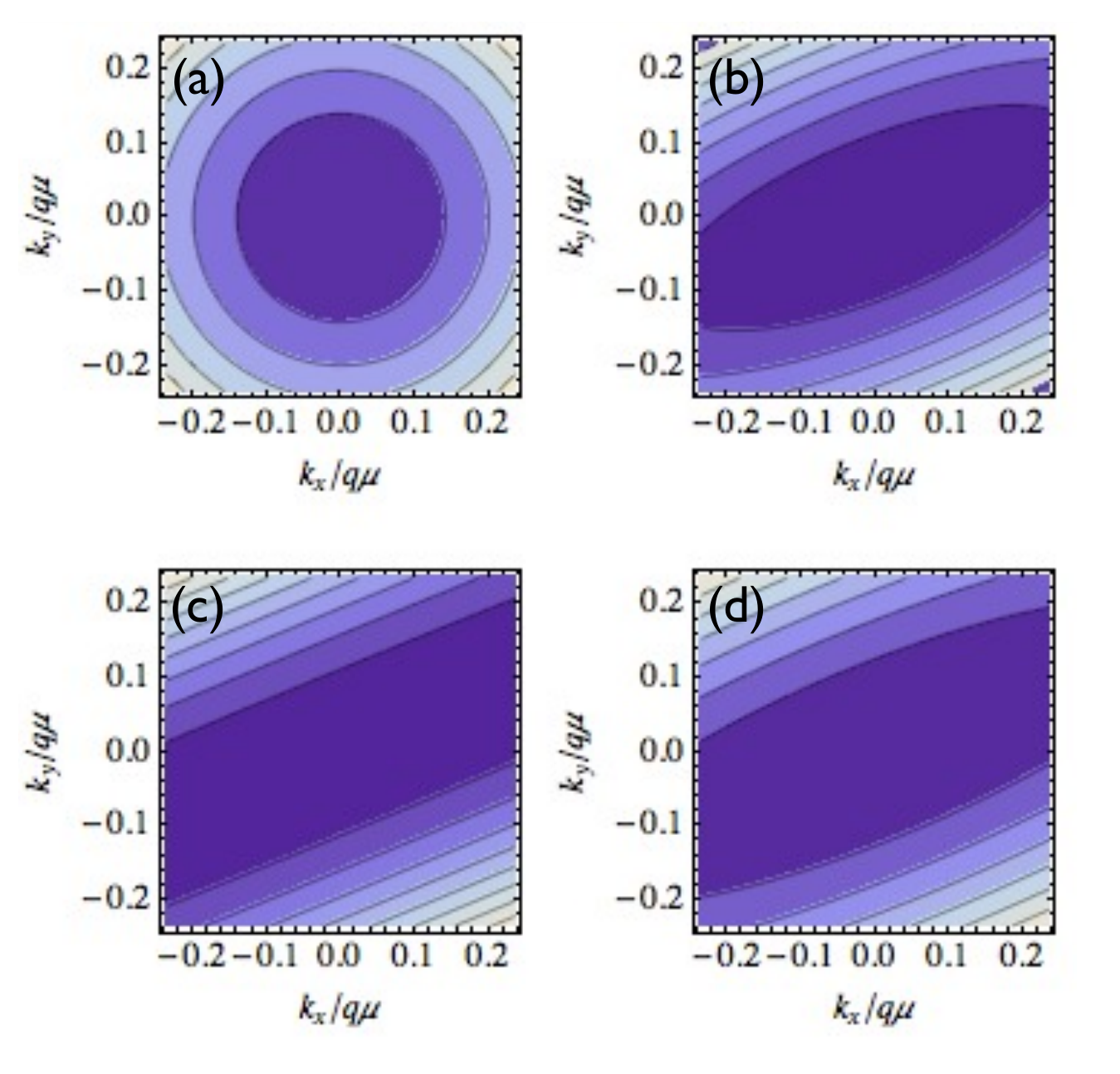}
\caption{\label{con}\footnotesize{Contour plot of $\tilde{k}_x'$ for a) $\lambda=0$, b) $\lambda=-0.4$, c) $\lambda=-0.95$ and d) $\lambda=-1.4$. The plots have been generated with $m=0$, $\omega=0$ and $q\mu=17$.} }
\end{figure}

In this section we employ a WKB approximation to uncloak uncloak the origin of the momentum-dependent spectral weight in Figure \ref{po}.   Such an approximation has been used recently in \cite{Hartnoll:2011dm} for the semi-analytic analysis of spectral functions in the electron star background \cite{Hartnoll:2010gu}.
Setting $k_x=k\cos\theta$ and $k_y=k\sin\theta$, we define new momentum coordinates 
\begin{align}\label{effk}
\tilde{k}^\prime_x&=\sqrt{\tilde{k}_x^2+\tilde{k}_y^2} \qquad \qquad \tilde{k}^\prime_y=0,
\end{align}
%
which are written in the so-called ``lab'' frame. Note that the expressions for $\tilde k_x$ and $\tilde k_y$ have been defined in \eqref{tildekx} and \eqref{tildeky}, respectively. With these new momenta and for a particular $\theta$ and $k$, the original Dirac equations \eqref{CoupledDiracEquation1} and \eqref{CoupledDiracEquation2} give rise to two decoupled equations
\begin{align}\label{DCoupledDiracEquation1}
\left({\tilde D}_r-i\sigma_2\sqrt{g^{xx}}\,{\tilde k}_x'\right)\psi_1&=0,\\
\left({\tilde D}_r+i\sigma_2\sqrt{g^{xx}}\,{\tilde k}_x'\right)\psi_2&=0.\label{DCoupledDiracEquation2}
\end{align}
Any anisotropy that arises now in the spectral function must be tied to $\tilde k_x^\prime$.  To confirm this, we construct a contour plot of  $\tilde{k}_x'$ at the horizon where the anisotropy is largest.  The plots in Figure \ref{con} show clearly that for a fixed value of $k$, $\tilde{k}_x'$ is minimum at $\theta=\pi/8$ and $9\pi/8$ which is consistent with  the values of $\theta$ where the spectral density is a minimum. 

Indeed, one can develop a better understanding on how $\tilde{k}_x'$ affects the spectral function by performing a WKB analysis of the Dirac equation. In order to make the notation less cluttered,  we will replace $\tilde{k}_x'$ by $k$ in what follows.  In order to apply the WKB approximation, we will take the limit where $k$, $m$, $\omega$ and $q$ are large, but their ratio remains constant. To this end, it is convenient to rescale all of these quantities with a factor $\gamma$ such that $k=\gamma \hat{k}$,
$m_\psi=\gamma \hat m$, $q=\gamma\hat q$  and $\omega=\gamma\hat\omega$, where $\gamma$ is taken to be large (compared to one).

\begin{figure}
\hskip -0.09in\includegraphics[width=6.0cm]{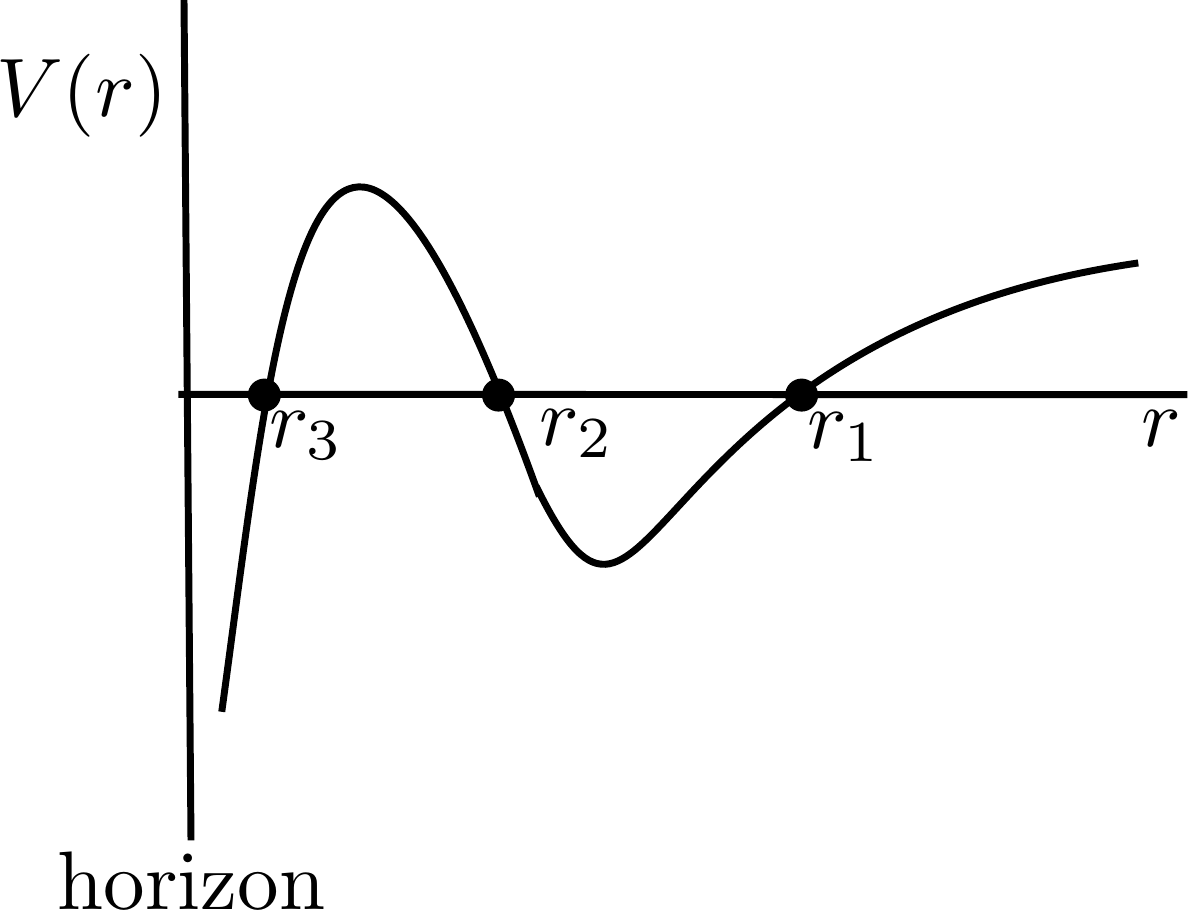}
\caption{\label{turn}\footnotesize{Potential $V$ in the Schr\"odinger equation as a function of the radial coordinate $r$.  The classical turning points as measured from the horizon are denoted by $r_3$, $r_2$, and $r_1$.}}
\end{figure}

Central to the WKB approximation is the construction of a Schr\"odinger-like equation for each of the components of the two-component spinor
\begin{align}
\psi_1=\left(\begin{array}{ll}\Phi_1\nonumber\\ \Phi_2\end{array}\right).
\end{align}
The same treatment applies equally to $\psi_2$ with the replacement of $k\to-k$.  To leading order in $\gamma$, the second-order differential equation for $\Phi_1$ takes on the form
\begin{align}\label{schr}
\Phi_1''=\gamma^2 g_{rr}\Big\{&\hat{m}^2+g^{tt}\left[\hat{\omega}+\mu \hat{q} \left(1-r_0/r\right)\right]^2\nonumber\\
&+g^{xx} \hat{k}^2\Big\}\Phi_1,
\end{align}
with
\begin{align}
\Phi_2=\frac{\left(-\sqrt{g^{rr}}\partial_r-m_\psi\right)\Phi_1}{-\sqrt{-g^{tt}}(\omega+q A_t)+\sqrt{g^{xx}} k}.
\end{align}
In the context of a Schr\"odinger equation, the right-hand side of equation \eqref{schr} can be regarded as the zero-energy potential
\begin{align}\label{SchrodingerPotential}
V=\gamma^2 g_{rr}\Big\{\hat{m}^2+g^{tt}\left[\hat{\omega}+\mu \hat{q} (1-r_0/r)\right]^2+g^{ii} \hat{k}^2\Big\}.
\end{align}
It is the turning points of this potential that governs the physics of the WKB approximation.
In the near-horizon region, the potential is approximated by
\begin{align}
V(r\to r_0)&=\frac{L^2\gamma^2}{3r_0^3(r-r_0)}\Big[L^2\hat{k}^2 +\hat{m}^2 r_0^2-\hat{\omega}^2\frac{L^2r_0}{3\left(r-r_0\right)}\Big]\nonumber\\
&+\cdots
\end{align}
If $\hat\omega$ is non-zero, the near horizon potential is dominated by the term proportional to $\hat\omega^2$ and is negative for real positive $\hat\omega$.
Close to the boundary, the leading order term of the potential
\begin{align}
V(r\to\infty)=\frac{\gamma^2L^2\hat{m}^2}{r^2}+\cdots,
\end{align}
is always positive.
Between these two regions, there can be at most three turning points which we denote by $r_1$, $r_2$ and $r_3$, with $r_1$ being the turning point closest to the boundary as depicted in Figure \ref{turn}.  For the parameter range of interest here, all three turning points will enter the matching conditions as we describe below.

What remains is a simple matching procedure to determine the form of $\Phi_1$ in each of the regions that bracket the turning points.  This procedure, detailed previously in \cite{Hartnoll:2011dm}, utilizes the functions
\begin{align}
X&=\int_{r_3}^{r_2} dr\sqrt{V}+\log2,\\
Y&=\int_{r_2}^{r_1}dr\sqrt{-V}+\frac{\pi}{2},
\end{align}
which directly enter the retarded Green function
\begin{align}
G_R(\omega,k)\propto\frac{i{\mathcal G}}{2}\lim_{r\to\infty}r^{2mL}{\rm exp}\left( -2\int_{r_1}^r dr' \sqrt{V}\right)
\end{align}
where
\begin{align}
{\mathcal G}=\frac{\cosh(X+iY)+\sinh(X-iY)}{\cosh(X-iY)-\sinh(X+iY)}.
\end{align}
The poles of the retarded Green function only come from $\mathcal G$ and are given by $Y=\pi n- ie^{-2X}+\cdots$ where we have included the imaginary part up to the leading order only. Since we are interested in the non-analytic dependence of the Green function on frequency and momentum, we will focus on the behavior in the vicinity of the pole where $\mathcal G$ can be written as
\begin{align}
{\mathcal G}=\sum_n\frac{i}{Y(\omega,k)-\pi n +i e^{-2X(\omega,k)}}.
\end{align}
Close to $\omega=0$ and $k=k_{\rm F}^{(n)}$, we can expand\footnote{Note that in equation \eqref{SchrodingerPotential}  we have written the potential in terms of the hatted quantities $\hat{\omega}$, $\hat{k}$, however, in doing the expansion of $Y(\omega,k)$, the factors of $\gamma$ cancel and we can drop the hat on $\omega$ and $k$.} $Y(\omega,k)$ as $Y\big(0,k^{(n)}_{\rm F}\big)+\omega\partial_\omega Y\big(0,k^{(n)}_{\rm F}\big) + (k-k_{\rm F}^{(n)})\partial_k Y\big(0,k^{(n)}_{\rm F}\big)+\cdots$.   As a result of this expansion, the retarded Green function takes the form
\begin{align}\label{GRfunc}
&G_R(\omega,k)\propto \sum_n \frac{ -c_n e^{-2 a_n}}{\omega+v_n(k-k_{\rm F}^{(n)})+i c_n e^{-2X(\omega,k)}}\\
&\propto\sum_n\frac{-\big[\omega+v_n(k-k_{\rm F}^{(n)})\big]c_ne^{-2a_n}+ic_n^2e^{-2a_n}e^{-2X}}{\big[\omega+v_n(k-k_{\rm F}^{(n)})\big]^2+c_n^2e^{-4X}},\nonumber
\end{align}
where
\begin{align}
v_n&=\frac{\partial_k Y\big(0,k^{(n)}_{\rm F}\big)}{\partial_\omega Y\big(0,k^{(n)}_{\rm F}\big)}, \nonumber\\
c_n&=\frac{1}{\partial_\omega Y\big(0,k^{(n)}_{\rm F}\big)}, \\
a_n&=\int^{\infty}_{r_1} dr \sqrt{V\big(0,k^{(n)}_{\rm F}\big)} ,\nonumber
\end{align}
which is essentially the equation (48) of \cite{Hartnoll:2011dm}.

Close to the Fermi surface, where the first term in the denominator of the second equation in \eqref{GRfunc}) can be dropped\footnote{We have assumed that in the limit $\omega\to0$, $\omega$ vanishes faster than $e^{-2X(\omega,k)}$ which is true in the range of parameters in which we are interested.}, the spectral function $A(\omega, k)$ can be extracted from the imaginary part of the retarded Green function. One then obtains
\begin{align}
A(\omega,k)\propto e^{-2a_n+2X}.
\end{align}
We can now understand the anisotropy of spectral function from the dependence of $a_n$ and $X$ on the area enclosed by the potential integrated between the turning points $r_2$ and $r_3$\footnote{This will define the quantity X.} and between $r_1$ and $\infty$ which will determine $a_n$.  In fact, for $m=0$, $r_1\rightarrow\infty$, $a_n$ therefore vanishes and the angular dependence of the spectral function will be dominated by $X$ alone.  Close to the horizon, the first turning point of the potential, $r_3$, is roughly given by
\begin{align}
r_3\approx r_0+\frac{\omega^2L^2r_0}{3\big[m^2r_0^2+k(r_0)^2L^2\big]}.
\end{align}
From this we can see that in the limit $\omega\to0$, that is, close to the Fermi surface, $r_3\to r_0$.

\begin{figure}
\hskip -0.09in\includegraphics[width=8.0cm]{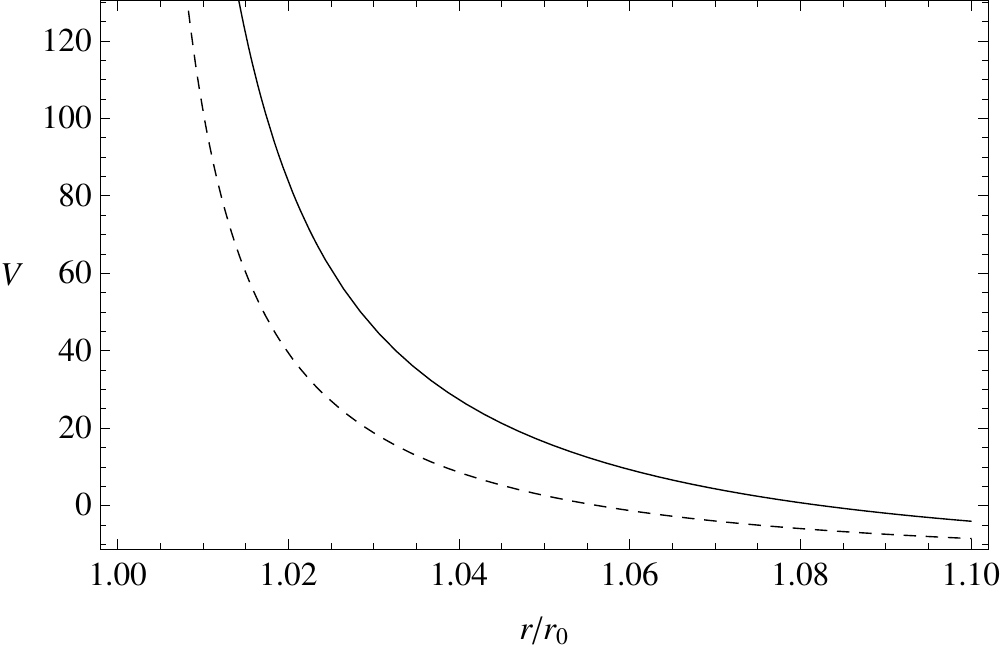}
\caption{\label{potential}\footnotesize{Effective WKB potential for $\lambda=-0.4$, $m=0$ at the first Fermi surface. The solid line corresponds to $\theta=0$ and the dashed line  to $\theta=\pi/8$.} }
\end{figure}

In our analysis of the spectral function, where we have set $m=0$, only $X$ determines the spectral function. To illustrate the angular dependence, we plot the near-horizon effective WKB Schrodinger potential for $\lambda=-0.4$, $m=0$ at first Fermi surface ($n=0$) in Figure \ref{potential} (corresponding to Figure 3(b)). The dashed line corresponds to the potential along $\theta=\pi/8$, while the solid line corresponds to $\theta=0$. It is now obvious that $r_2$ is larger for $\theta=0$ than for $\theta=\pi/8$ and the potential is always larger for $\theta=0$ than $\theta=\pi/8$ in the range of $r$ shown. Hence $X$ and spectral weight are larger along $\theta=0$ than $\theta=\pi/8$, thereby explaining our key finding that the spectral weight is angularly dependent.  In principle, the WKB approximation works only for large $n$, nevertheless, we see that the result qualitatively agrees with our numerical results even for small $n$. 

\section{Discussion}

We have shown here how holography can be used to model a shape
distortion of the underlying Fermi surface in a non-Fermi liquid through the condensation of a neutral symmetric traceless operator which is dual to a neutral massive spin-two field in the bulk.  An open challenging problem remains: Is the condensation of the boundary theory operator ${\cal O}_{ij}$ possible starting from a bulk geometry which is not an Einstein manifold?  In other words, what is the backreacted background that allows for the ghost-free and causal propagation of a neutral massive spin-two field. These extensions are extremely desirable since they would not only enable an analysis at zero-temperature but would also cure at finite temperature some of the peculiar thermodynamical properties of the conduced phase in the probe limit. In addition, is the angular dependent spectral weight  a generic feature of holographic non-Fermi liquids with (partially) broken rotational symmetry?  If so, then holography would have provided a key ingredient missing from most condensed matter analyses of the Pomeranchuk instability, namely a vanishing spectral weight for a finite range of momenta and hence a pseudogap as a function of frequency. Also, how does the resistivity tensor behave in the two different directions? In our probe analysis, where the backreaction of  the massive spin-two field on the metric is ignored, the resistivity tensor is not sensitive to the anisotropy of the system when the boundary theory is in the broken phase.   Nevertheless, one might be able to see the anisotropic contribution of the fermions to the resistivity through a fermion loop computation in the bulk similar to the analysis of  \cite{Faulkner:2010da}.

\acknowledgements 
We would like to thank R. Leigh, C. Herzog, and E. Fradkin for several
important discussions at the outset of this work. P.  W. P. would like to thank S. Hartnoll for
comments on the manuscript.  M. E. is supported by the NSF under Grant Number PHY-0969020 while K.  L. and P. W. P.  acknowledge financial support from the NSF-DMR-0940992 and DMR-1104909.

\section*{Appendix}

In this appendix, we demonstrate that adding a coupling of the form $\ell^2C_{\gamma\delta\rho\sigma}C^{\gamma\delta\rho\sigma}(\varphi_{\mu\nu}\varphi^{\mu\nu}-\varphi^2)$ to the BGP Lagrangian, given that the background spacetime is an Einstein manifold, does not induce any extra degrees of freedom for the neutral massive spin-two field $\varphi_{\mu\nu}$ nor violate its causal propagation. 
Our discussion below closely follows that of \cite{Buchbinder:2000fy}, where it has been shown that neutral massive spin-two field in an Einstein spacetime, such as the AdS$_{d+1}$ Schwarzschild black hole, has the correct number of degrees of freedom and propagates causally. In our discussion below, we keep the number of dimensions of the bulk spacetime (denoted by $d+1$) arbitrary. 

Consider the following Lagrangian \cite{Aragone:1971kh}
\begin{align}\label{ADAction}
{\cal L}=\frac{1}{4}&\Big\{\hskip-0.02in-\nabla_\mu \v_{\nu\rho} \nabla^\mu\v^{\nu\r} +\nabla_\mu\v\nabla^\mu\v-2\nabla^\mu\varphi_{\mu\nu}\nabla^\nu\v\nonumber\\
& +2\nabla_\mu \v_{\nu\r} \nabla^\r \v^{\nu\mu} 
-m_0^2(\varphi_{\mu\nu}\varphi^{\mu\nu}-\varphi^2)\nonumber\\
&+2a_1 R\,\varphi_{\mu\nu}\varphi^{\mu\nu}+2a_2 R\,\varphi^2
+2a_3 R^{\mu\lambda\nu\rho}\varphi_{\mu\nu}\varphi_{\l\r} \nonumber\\
&+2a_4 R^{\mu\nu}\varphi_{\mu\l}\varphi_\nu^\l+2a_5 R^{\mu\nu}\varphi_{\mu\nu}\varphi \Big\},
\end{align}  
where $\varphi=\varphi_{\,\mu}^\mu$. The above Lagrangian is the most general two-derivative action for a neutral massive spin-two field $\varphi_{\mu\nu}$ (up to the quadratic order in $\varphi_{\mu\nu}$, though) in a curved background. The coefficients in the first two rows of \eqref{ADAction} are fixed because in the flat spacetime limit, one demands the Lagrangian to go over to the Fierz-Pauli Lagrangian \cite{FP1939}, which is known to have the correct number of causally-propagating degrees of freedom for a neutral massive spin-two field. Indeed, in order for a  massive spin-two field (represented by a symmetric tensor $\varphi_{\mu\nu}$) to have the correct number of propagating degrees of freedom in a $(d+1)$-dimensional spacetime, one needs $2(d+2)$ constraint equations to kill the extra degrees of freedom.

Now, add to the above Lagrangian an additional coupling of the form $\ell^2C^2(\varphi_{\mu\nu}\varphi^{\mu\nu}-\varphi^2)$ and define the Lagrangian
\begin{align}
{\cal L}_{\varphi}={\mathcal L}+\frac{1}{4}\ell^2C^2\left[\v_{\mu\nu}\v^{\mu\nu}-
\v^2\right], 
\end{align}
where, in order to make the expressions less cluttered, we defined $C^2=C_{\gamma\delta\rho\sigma}C^{\gamma\delta\rho\sigma}$. For $\ell=0$, the analysis has been performed in \cite{Buchbinder:2000fy}, where, in order to get rid of the extra degrees of freedom of $\varphi_{\mu\nu}$, one obtains a one-parameter family of solutions for the coefficients $a_1, \cdots, a_5$ (labeled by an arbitrary real parameter $\xi$) which can be represented as 
\begin{align}\label{as}
a_1=\frac{\xi}{d+1}, \quad a_2=\frac{1-2\xi}{2d+2}, \quad a_3=a_4=a_5=0,
\end{align}
provided that the background metric $g_{\mu\nu}$ satisfies the Einstein relation $R_{\mu\nu}=2\Lambda g_{\mu\nu}/(d-1)$, and $m_0^2\neq 2(\xi-1)/(d+1)$. Defining $m^2=m_0^2+2(1-\xi)R/(d+1)$ and given the above solution for $a_i$'s, the Lagrangian \eqref{ADAction} takes the form given in the expression \eqref{BGPAction}, the so-called BGP Lagrangian.

For $\ell\neq 0$, one can easily obtain the constraint equations following \cite{Buchbinder:2000fy}. Basically, constraints come from the equations of motion and their derivatives which do not contain two time derivatives of spin-two field $\ddot{\v}_{\mu\nu}$. The equations of motion read
\begin{align}\label{TildeE}
\tilde E_{\mu\nu}=E_{\mu\nu}+\ell^2C^2\left(\v_{\mu\nu}-g_{\mu\nu}\varphi\right)=0,
\end{align}
where $E_{\mu\nu}$ denotes the equations of motion for the $\ell=0$ case. The explicit form of $E_{\mu\nu}$ is given in \cite{Buchbinder:2000fy}. Since $E_{\mu0}$ does not involve second time-derivatives of $\varphi_{\mu\nu}$, and obviously the new term in \eqref{TildeE} does not involve any second time-derivative of $\varphi_{\mu\nu}$ either,  the equations $\tilde E_{\mu0} = 0$ actually give us $d+1$ (primary) constraints. 

In order to obtain the secondary constraints, we take the covariant derivative of the equations of motion
\begin{align}\label{con2}
\nabla^\mu \tilde E_{\mu\nu}=0.
\end{align}
Again, neither $\nabla^\mu E_{\mu\nu}$ nor $\nabla^\mu \left[C^2(\varphi_{\mu\nu}\varphi^{\mu\nu}-\varphi^2)\right]$ involves two time-derivatives of the spin-two field. Thus, the equations \eqref{con2} give us $d+1$ secondary constraints. Following \cite{Buchbinder:2000fy}, the equations \eqref{con2} should contain $\dot{\v}_{\mu0}$ through a rank $d$ matrix in order for their derivatives to define $d$ accelerations $\ddot{\v}_{i0}$:
\begin{align}
{E_{\mu 0,}}^\mu&={\cal A}\,\dot{\v_{00}}+{\cal B}^j\dot{\v_{j0}}+\cdots,\nonumber\\
{E_{\mu i,}}^\mu&={\cal C}_i\,\dot{\v_{00}}+{\cal D}^j_i\dot{\v_{j0}}+\cdots,\\
{\rm rank}\,&\hat{\Phi}_\mu^\nu\equiv {\rm rank} \left\| \begin{array}{cc}
{\cal A} & {\cal B}^j \\
{\cal C}_i & {\cal D}_i^j \\
\end{array} \right\|=d.\nonumber
\end{align}
The explicit form for $\hat{\Phi}_\mu^\nu$ when $\ell=0$ appears in \cite{Buchbinder:2000fy}. The additional non-vanishing contributions to ${\cal A}$, ${\cal B}^j$, ${\cal C}_i$ and ${\cal D}^i_j$ are
given by
\begin{align}
&\delta {\cal B}^j=-\ell^2 C^2 g^{j0},\qquad \delta {\cal D}^i_j=\ell^2 C^2 g^{00} \delta^i_j.
\end{align}
Hence, in order for the rank of $\hat{\Phi}^\nu_\mu$ to be $d$, we require $m_0^2-\ell^2C^2+2(1-\xi)R/(d+1) \neq0$, where $\xi$ is arbitary parameter. We now have $2d+2$ constraints. 

One of the remaining constraints can be obtained from a linear combination of the equations of motion and the primary and secondary constraints:
\begin{align}\label{con3}
&\frac{m_0^2-\ell^2 C^2}
{d-1}\tilde E_{\mu}^{\mu}+\nabla^\mu\nabla^\nu \tilde E_{\mu\nu}+\frac{2(1-\xi)}{(d+1)(d-1)}R \tilde E_{\mu}^{\mu}\nonumber\\
&=\frac{\v}{d-1}\left[\frac{2(1-\xi)}{d+1}R+m_0^2-\ell^2 C^2\right]\nonumber\\
&\times\left[\frac{d+1-2\xi d}{d+1}R+(m_0^2-\ell^2 C^2)d\right]\\
&+ 2 \ell^2(\nabla^\nu C^2)(\nabla^\mu \v_{\mu\nu}-\nabla_\nu \v)\nonumber\\
&+\ell^2(\nabla^\mu\nabla^\nu C^2)(\v_{\mu\nu}-g_{\mu\nu}\v)
\approx 0.\nonumber
\end{align}
This equation can be used as a constraint since it does not contain $\ddot{\v}_{\mu\nu}$. It contains $\dot{\v}_{\mu\nu}$ which can be removed by the equations $\nabla^\mu \tilde E_{\mu\nu}=0$. After such a removal, the time-derivative of \eqref{con3} does not contain $\ddot{\v}_{\mu\nu}$ and can be used as the last constraint. We then have in total $2d+4$ constraints and hence the correct number of propagating degrees of freedom. 

Following the discussion in \cite{Buchbinder:2000fy}, it can be seen easily that the characteristic matrix  remains unchanged for $\ell\neq 0$, due to the fact that the Weyl squared coupling does not affect any of the terms with two time-derivatives. So,  the equations remain hyperbolic and causal even for the $\ell\neq 0$ case. The resulting equations of motion are
\begin{align}
\left(\nabla^2-m^2+\ell^2C^2\right)\v_{\mu\nu}+2R^{\rho\,\,\,\sigma}_{\,\,\,\mu\,\,\,\nu}\varphi_{\rho\sigma}=0.
\end{align}
where, as we have defined before, $m^2=m_0^2+2(1-\xi)R/(d+1)$.

\end{document}